# A high-mass X-ray binary descended from an ultra-stripped supernova

Noel D. Richardson[1], Clarissa Pavao[1], Jan J. Eldridge[2], Herbert Pablo[3], André-Nicolas Chené[4], Peter Wysocki[5], Douglas R. Gies[5], George Younes[6,7], Jeremy Hare[6]

[1] Department of Physics and Astronomy, Embry-Riddle Aeronautical University, 3700 Willow Creek Rd, Prescott, AZ, 86301, USA; noel.richardson@erau.edu

[2] Department of Physics, University of Auckland, Private Bag 92019, Auckland, New Zealand

[3] American Association of Variable Star Observers, 49 Bay State Road, Cambridge, MA 02138, USA

[4] Gemini Observatory, Northern Operations Center, 670 A'ohoku Place, Hilo, HI 96720, USA

[5] CHARA and the Department of Physics and Astronomy, Georgia State University, Atlanta, GA 30302-5060, USA

[6] NASA Goddard Space Flight Center, Greenbelt, MD 20771, USA

[7] Department of Physics, The George Washington University, Washington, DC 20052, USA

**Ultra-stripped supernovae are different from other terminal explosions of massive stars, as they show little or no ejecta from the actual supernova event [1,2]. They are thought to occur in massive binary systems after the exploding star has lost its surface through interactions with its companion [2]. Such supernovae produce little to no kick, leading to the formation of a neutron star without loss of the binary companion, which itself may also evolve into another neutron star [2]. Here we show that a recently discovered high-mass X-ray binary, CPD -29 2176 (CD -29 5159; SGR 0755-2933) [3,4,5,6], has an evolutionary**

**history that shows the neutron star component formed during an ultra-stripped supernova. The binary has orbital elements that are similar both in period and in eccentricity to one of 14 Be X-Ray binaries that have both known orbital periods and eccentricities [7]. The identification of the progenitors systems for ultra-stripped supernovae is necessary as their evolution pathways leads to the formation of a binary neutron star systems. Binary neutron stars, such as the system that produced the kilonova GW170817 that was observed with both electromagnetic and gravitational energy [8], are known to produce a large quantity of heavy elements [9,10].**

The Be star CPD -29 2176 was found coincident on the sky to SGR 0755-2933 (2SXPS J075542.5−293353) after the object experienced a magnetar-like burst [3,4,5] (see Methods section for details). Ground-based optical spectroscopy of the object was carried out to analyze the motion of the Be star and prove the Be star was bound to the neutron star. The star has been classified as both B0Ve [11,12,13] and a hotter O6/8 star [14]. Our ground-based spectroscopy only shows the one He II line at 4686Å, consistent with a B0Ve star as an O star classification would have more He II lines. The spectrum is also typical of a Be star, showing emission from hydrogen Balmer lines, singly-ionized iron, and neutral helium. Our data only show the high excitation line of ionized helium at 4686Å in absorption, representing the only emission-free photospheric absorption line in the optical. Radial velocities of this line (Extended Data Table 1) show a periodic signal with a period of 59.5 d and a semi-amplitude of 2.76 km s$^{-1}$ (Table 1). Our orbit fits show this to be a circular orbit when considering the residuals for both it and a slightly eccentric binary [15], although we provide both solutions in Table 1. This represents a single-lined spectroscopic orbit for the system, which is shown in Figure 1. For any spectroscopic orbit, the measured mass function relates the masses of the two component stars to the inclination of the system. This function is dependent on the eccentricity $e$, the semi-amplitude of the orbit, and the period $P$ of the orbit in d. We show this relationship for a range of inclinations in Figure 1. If we assume that the companion is a neutron star with the Chandrasekhar mass of 1.44 $M_\odot$, then the allowed inclination for a normal mass of a B0 V star implies a nearly face-on inclination between ~10° and ~15°. This is fully consistent with the observed emission lines in the system which are narrow for Be stars, and like other nearly face-on classical Be stars.

The orbital separation is an order of magnitude greater than the expected Be star radius. Additionally, the mass ratio between the neutron star and Be star is below 0.1. These factors by a simple estimate suggest the circularization timescale for the binary is of the order of 100 Myrs [16], which is longer than the age of the system. Of note, the line-of-sight velocity ($\gamma$) is measured to be 58.1 km/s, which is large for nearby stars. Such a high velocity is often associated with runaway stars, where these stars have managed to obtain a velocity large enough to leave their original orbits around the Galactic Plane due to either multi-body interactions or supernovae kicks. However, CPD -29 2176 is not nearby, but instead in the outer Galactic Plane, where such velocities are expected and do not indicate runaway status. As such, it probably has a near-circular orbit in the Galaxy, which implies that the formation of the neutron star did not impart a significant velocity kick to the system. It has been shown that the Be X-Ray binaries often show typical kinematics in the Galactic potential [17], making CPD -29 2176 a normal system with respect to its $\gamma$ -velocity.

To understand the previous and future evolution of this binary system, we have searched through the Binary Population and Spectral Synthesis (BPASS) v2.2 binary evolution models calculated with Solar metallicity (Z=0.02) [18,19]. We first searched for models that match the system as currently observed. To do this, we selected only the BPASS secondary models where a normal star orbits a compact companion. We require that $\log(T_{eff}/K) = 4.44\pm0.05$, the orbital period is 59.5 days and that the star is on the main sequence. We constrain the luminosity of the star by assuming a distance 3.84 kpc for the system and use *UBVJHK* magnitudes taken from SIMBAD. We estimate extinction due to dust of $A_V = 1.8$ mag according to fits of the stellar spectral energy distribution.

This search identifies several models that match. For the next step, we now consider the evolution that occurs prior to the formation of the compact companion. Here we search for the BPASS primary models which give rise to our selected secondary models. To reduce the list of possible models further we consider what happens at the first supernova. Here we use the constraint that the binary is today circular, this requires us to pick out the models that have the smallest ejecta mass with $M_{ejecta} < 0.1 M_{total}$ so that the orbit remains circular, and the systemic velocity remains within the normal range for massive stars in the Galactic plane. By applying this filter, we only find two models for which the ejecta masses are sufficiently small, indicating a high likelihood of being observed today with nearly circular orbits. The parameters of the two found binary models are given in Table 2.

We show the evolution of these models in Figure 2. We see in the HR diagram that mass transfer occurs towards the end of the main sequence of the primary star. This leads to efficient mass transfer, substantially increasing the secondary mass and reducing the primary to a low-mass helium star. The primary then collapses as a low-mass helium star to form a neutron star. The system is today observed only a few million years after the supernova. In the future, the secondary will interact with the neutron star, and the secondary will become a stripped helium star towards the end of its lifetime. The source for the low kick from the in the core-collapse could be related to the nature of the supernova. In the models we present the final CO core masses are about 1.5 $M_\odot$. With this mass, and because it lost their hydrogen envelopes, the progenitor will not experience an electron-capture supernova in an ONe core. They are likely to proceed towards an iron-core collapse [20]. Electron-capture progenitors could be possible but would require lower mass progenitors or a different mass transfer history to obtain a lower core mass for the progenitor [21]. However, the nature of the core-collapse is not important as the ejecta mass and the resultant kick is always the limiting constraint on providing the nearly circular orbit. Further study of the neutron star in this system may lead insight into the nature of the collapse that led to its formation.

In Figure 2, we also show how the masses of the two stars and the radii vary with age. We see that there are two mass transfer events for the primary. The first event occurs at the end of its main sequence and a second episode occurs a few million years after the primary star attempts to become a cool supergiant. The star also approaches a final interaction when it grows in radius to become a helium giant. We note that in all cases the mass transfer is stable, and the system never experiences common-envelope evolution. Given that the final primary mass is close to 2 $M_\odot$ and the total mass of the system is close to 20 $M_\odot$, we expect that the orbit will be minimally affected by the core collapse of the original primary component. We see that after the first supernova the system will be observable in its current state for many millions of years.

There is of course some uncertainty in binary evolution, but we believe that this picture is robust with the biggest uncertainty being whether the original primary star experiences a supernova as an ultra-stripped supernova or if it directly collapsed into the neutron star. However, given that the companion is so massive, the mechanics of the core-collapse have little impact on the orbit that we observe today. Also given that none of the interactions are common-envelope evolution but all stable mass transfer, the remaining uncertainties are related to how much systemic mass loss occurs in the interactions (non-conservative evolution). The future evolution of the system will see the size of the orbit shrink as mass is transferred from the Be star toward the neutron star and then ejected from the system. The current-day Be star should then explode leaving behind a neutron star to create a binary neutron star system. With the BPASS models, we find the system will last in its present state for 0.9 Myr with an expected 3.3 such systems per $10^6$ $M_\odot$ of stars. Thus, if we assume the current star formation rate in the Milky Way is 3.5 $M_\odot$ yr$^{-1}$, then we would expect about ten such systems to be found at present in the Galaxy. This system reveals that some neutron stars are formed with only a small supernova kick. As we understand the growing population of similar systems [22], we will gain insight into how calm some stellar deaths may be and if these stars can die without supernovae [23].

**Correspondence**

Correspondence and requests for materials should be addressed to the corresponding author Noel D. Richardson (noel.richardson@erau.edu).

**Acknowledgements**

CMP acknowledges support from the Embry-Riddle Aeronautical University's Undergraduate Research Institute and the Arizona Space Grant. This research was partially supported through Embry-Riddle Aeronautical University's Faculty Innovative Research in Science and Technology (FIRST) Program. The spectroscopy from CTIO was collected through NOIR Lab programs 2018B-0137 and 2020A-0054.

**Author Contributions**

The project was started and spectroscopy was proposed for by H.P., N.D.R., and A.-N.C. J.H. and G.Y. confirmed the astrometry of the neutron star with Swift and Chandra observations, showing it coincident with the Be star. C.P. reduced and analyzed the spectroscopic data with guidance from N.D.R. The Galactic kinematics were done by P.W. and D.R.G. J.J.E. modeled the binary evolution of the system. All authors discussed and commented on the manuscript.

**Competing interests**

The authors declare no competing interests.

**Data availability**

The reduced spectroscopic data that support the plots within this paper and other findings of this study are available from the corresponding author upon request. The raw data are available from the NOIR Lab archive. BPASS results and stellar models are available from bpass.auckland.ac.nz.

**Code availability**

The data analysis code used in this analysis is all open-source software. BPASS results and stellar models are available from bpass.auckland.ac.nz.

**References**

[1] De, K. et al. A hot and fast ultra-stripped supernova that likely formed a compact neutron star binary. Science 362, 201 (2018)

[2] Tauris, T. M., Langer, N., & Podsiadlowski, P. Ultra-stripped supernovae: progenitors and fate. Mon. Not. R. Astron. Soc. 451, 2123 (2015).

[3] Barthelmy, S. D. et al. Swift detection of a likely new SGR: SGR 0755-2933. The Astronomer's Telegram 8831, 1 (2016).

[4] Archibald, R. F. et al. Swift XRT Observations of SGR J0755-2933. The Astronomer's Telegram 8868, 1 (2016).

[5] Surnis, M. P. et al. Upper limits on the pulsed radio emission from SGR candidate SGR 0755-2933. The Astronomer's Telegram 8943, 1 (2016)

[6] Doroshenko, V., Santangelo, A., Tsygankov, S. S., & Ji, L. "SGR 0755-2933: a new high-mass X-ray binary with the wrong name". Astron. Astrophys. 647, 165 (2021).

[7] Reig, P. "Be/X-ray binaries". Astrophysics and Space Science, 332, 1 (2011).

[8] Abbott, B. P. et al. "Multi-messenger Observations of a Binary Neutron Star Merger". Astrophys. J.. Letters. 848, L12 (2017).

[9] Chornock, R., et al. "The Electromagnetic Counterpart of the Binary Neutron Star Merger LIGO/Virgo GW170817. IV. Detection of Near-infrared Signatures of r-process Nucleosynthesis with Gemini-South", Astrophys. J.. Letters. 848, L19 (2017).

[10] Watson, D., et al. "Identification of strontium in the merger of two neutron stars", Nature, 574, 497 (2019).

[11] Fernie, J. D., Hiltner, W. A., & Kraft, R. P. "Association II PUP and the classical Cepheid AQ Pup.", Astronom. J., 71, 999 (1966).

[12] Reed, B. C., & Fitzgerald, M. P. "A photoelectric UBV catalogue of 610 stars in Puppis", Mon. Not. R. Astron. Soc. 205, 241 (1983).

[13] Vijapurkar, J., & Drilling, J. S. "MK Spectral Types for OB + Stars in the Southern Milky Way", Astrophys. J. Supp., 89, 293 (1993)

[14] Reed, B. C. "Catalog of Galactic OB Stars", Astronom. J, 125, 2531 (2003)

[15] Lucy, L. B. "Spectroscopic binaries with elliptical orbits". Astron. Astrophys. 439, 663 (2005).

[16] Eldridge J. J., Tout C. A., 2019, The Structure and Evolution of Stars. World Scientific, doi:10.1142/p974

[17] Chevalier, Claude and Ilovaisky, Sergio A. "HIPPARCOS results on massive X-ray binaries". Astron. Astrophys. 330, 201 (1998).

[18] Eldridge, J. J. et al. "Binary Population and Spectral Synthesis Version 2.1: Construction, Observational Verification, and New Results". Pub. Astron. Soc. Aus. 34, 58 (2017).

[19] Stanway, E. R., & Eldridge, J. J. "Re-evaluating old stellar populations". Mon. Not. R. Astron. Soc. 479, 75 (2018).

[20] Woosley, S. E., & Heger, A. "The Remarkable Deaths of 9-11 Solar Mass Stars". Astrophys. J., 810, 34 (2015).

[21] Podsiadlowski, Ph., et al. "The Effects of Binary Evolution on the Dynamics of Core Collapse and Neutron Star Kicks". Astrophys. J., 612, 1044 (2004).

[22] Shenar, T. et al. "An X-ray quiet black hole born with a negligible kick in a massive binary within the Large Magellanic Cloud" Nature Astronomy. 6, 1085 (2022).

[23] Allan, A. P. et al. "The possible disappearance of a massive star in the low-metallicity galaxy PHL 293B" ". Mon. Not. R. Astron. Soc. 496, 1902 (2018).

Table 1: Orbital Elements of CPD -29 2176.

| Orbital Element | Value (Circular solution) | Value (Elliptical solution) |
|---|---|---|
| Period $P$ | 59.52 ± 0.55 d | 59.69 ± 0.15 d |
| Eccentricity $e$ | 0.0 (fixed) | 0.06 ± 0.06 |
| $\omega$ | 0.0° (fixed) | 165.9° ± 4.2° |
| Semi-Amplitude $K_1$ | 2.76 ± 0.49 km s$^{-1}$ | 2.16 ± 0.15 km s$^{-1}$ |
| Line of sight velocity $\gamma$ | 58.07 ± 0.35 km s$^{-1}$ | 58.62 ± 0.11 km s$^{-1}$ |
| Time of periastron | HJD 2458387.64 ± 2.75 | HJD 2458353.24 ± 0.70 |
| r.m.s. of residuals | 1.65 km s$^{-1}$ | 1.74 km s$^{-1}$ |

Table 2: Binary Model Parameters for evolutionary history discussed in the text.

| Parameter | Model 1 | Model 2 |
|---|---|---|
| $M_1$;initial | 12 $M_\odot$ | 12 $M_\odot$ |
| $M_2$;initial | 9.6 $M_\odot$ | 8.4 $M_\odot$ |
| log(P=days) [initial] | 0.4 | 0.4 |
| $M_1$, remnant | 1.4 $M_\odot$ | 1.4 $M_\odot$ |
| $M_2$, post-SN | 19 $M_\odot$ | 18 $M_\odot$ |
| log(P=days) [post-SN] | 1.8 | 1.8 |

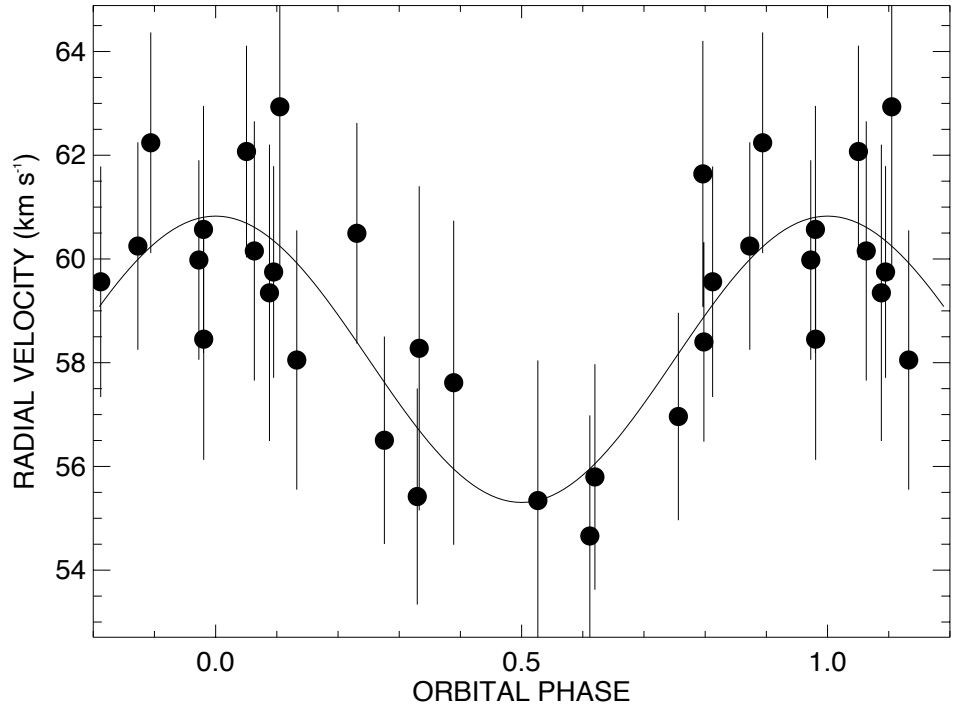

**Figure 1.** Orbit and Orbital Properties of CPD -29 2176. In Panel (a), we show the single-lined spectroscopic orbit of CPD -29 2176 with the circular orbital elements given in Table 1 with 1σ error bars for the data. Panel (b) shows the mass function of the binary with a nominal range of masses for the Be star (vertical dashed lines) and the Chandrasekhar mass shown as a horizontal dashed line. The mass relations for a variety of inclinations are shown.

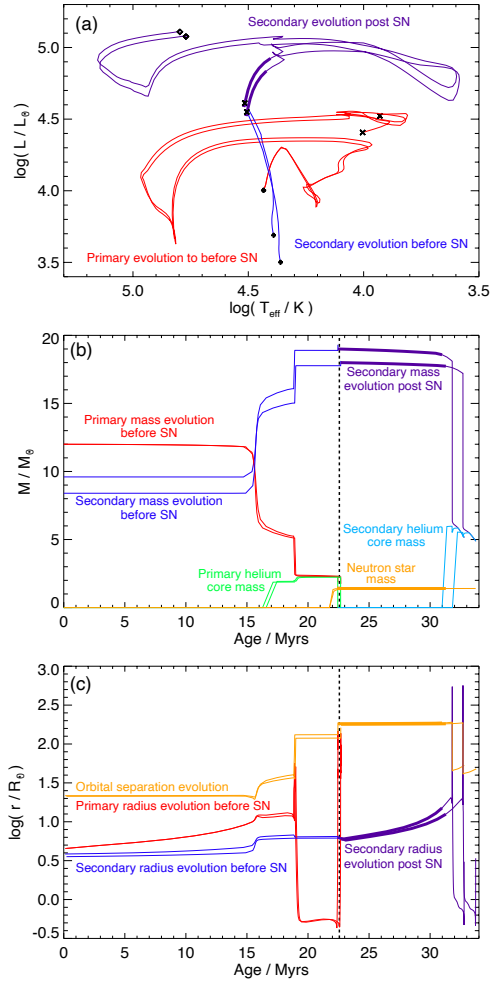

**Figure 2.** Evolution of the binary system. In panel (a), we show the Hertzsprung-Russell diagram, in red is the evolution of the primary star, in blue the evolution of the secondary star during the primary evolution and in purple it's evolution in the detailed secondary model after the primary star's supernova. Panel (b) shows the mass evolution of the binary system. Again, the colours are the same as in the left panel but here green shows the evolution of the primary star's helium core and the orange line the mass of the CO core and then the neutron star remnant. In panel (c), we show the stellar radii (in blue and red) and the orbital radius in orange. In both panels (b), and (c), we indicate the time of the primary star's supernova event as a vertical dashed line.

# Methods

**Chandra X-ray Observatory**

The Chandra X-ray Observatory observed SGR 0755-2933 using the Advanced CCD Spectrometer (ACIS) [24] on 2019 October 3 (MJD 58759.2; obsID 22454) for 29.69 ks. The source was imaged on the front-illuminated ACIS-I3 chip in timed exposure mode and the data were telemetered using the "FAINT" format. The data were reduced and analyzed using the Chandra Interactive Analysis of Observations (CIAO) software version 4.12 with CALDB version 4.9.2.1. Prior to analysis, the Chandra event file was reprocessed using the CIAO tool chandra_repro.

The CIAO tool wavdetect was run on the 0.5-8 keV energy band image to locate all sources in the field of view. In total, 31 sources were detected at a signal-to-noise ratio (S/N) larger than 3. SGR 0755-2933 was by far the brightest source detected in the field with a S/N of 756. The 90% absolute astrometric accuracy of Chandra is 0.8" [25]. However, this can be partially corrected for by cross-matching X-ray sources with Gaia optical sources and correcting Chandra's absolute astrometry [26]. We found four X-ray sources (excluding SGR 0755-2933), located within 6' of the aim point and detected at a S/N> 5, with Gaia optical counterparts located within 1" of the X-ray source position. Prior to any astrometric correction, the average offset between the four Chandra sources and their Gaia counterparts was 0.56". We then used the CIAO tools wcs_match, to calculate an updated astrometric solution, and wcs_update to apply the coordinate transform to the event file. An offset of 0.54" ($\Delta\alpha$=-0:47", $\Delta\delta$= -0:27") was found and corrected for. After applying the correction, the average offset between the four Chandra sources and their Gaia counterparts improved to 0.18". We adopt this average residual offset as the remaining systematic 1σ uncertainty on the absolute astrometry.

After correcting the astrometry we reran the CIAO tool wavdetect on the 0.5-8 keV energy filtered image. SGR 0755-2933 is detected at a position of R.A.= 118:926997(3)°, Decl.= -29:564849(3)°. We calculate the source's positional uncertainty by adding the statistical uncertainties of the R.A. and Decl. to the systematic residual uncertainties from the absolute astrometry in quadrature and find the positional uncertainty of the source to be 0.4" at the 2σ level.

**Multi-wavelength Counterpart**

There is a single potential multi-wavelength counterpart located within 1" of the Chandra source position of SGR 0755-2933. This source has a Gaia position offset from the X-ray position of only 0.09". The Gaia measured parallax, $\pi = 0.26 \pm 0.03$ mas, and proper motion, $\mu_\alpha \cos\delta = -2.66 \pm 0.04$ mas yr$^{-1}$, $\mu_\delta = 2.68 \pm 0.05$ mas yr$^{-1}$ [27], of the source place it at a distance of $3.36 \pm 0.16$ kpc based on the inference method that has been implemented [28]. The source is relatively bright having a Gaia G-band magnitude, $G$=9.94, and has been classified as a B0 Ve type star through optical spectroscopy [11,12,13]. The Be type stars typically have a ``decretion" disk due to their fast rotation [29]. Additionally, the source flux has also been observed to be variable [30,31].

Given the neutron star nature of SGR 0755-2933, it is important to understand whether or not this potential optical counterpart could be a binary companion to the neutron star or just be a chance coincidence overlap. To calculate the chance coincidence probability of the neutron star and Be star being spatially collocated, we calculate the probability of one or more Gaia sources landing within the 2σ X-ray positional uncertainty of SGR 0755-2933 (i.e., $\Delta r^2_{2\sigma}$) by chance. This probability is $P_{Gaia} = 1 - \exp(-\rho_{Gaia}\pi\Delta r^2_{2\sigma})$, where $\rho_{Gaia}$ is the density of Gaia sources within a 5' radius surrounding SGR 0755-2933. Using the Gaia source density of $7\times10^{-3}$ arcsec$^{-2}$, we find a chance coincidence probability of $P_{Gaia}$ ~0.4%. If instead we only consider bright optical sources (i.e., $G<10$), which are relatively rare, the chance coincidence probability drops to $P_{Gaia}$ ~$4\times10^{-4}$%. Therefore, it is unlikely that these sources are collocated by chance, and much more likely that bright optical source is the binary companion to SGR 0755-2933.

**Radial Velocities and a Spectroscopic Orbit**

Our optical spectroscopy data were collected with the Cerro Tololo Interamerican Observatory's 1.5 m telescope and the CHIRON spectrograph [32]. The data were reduced with the standard pipeline for reductions that was recently described in the literature [33]. We extracted the order containing the He II λ4686 absorption profile, and we fit the line with a Gaussian profile to measure the centroid position. An example fit is shown in Figure Extended Data Figure 1, and our measurements are tabulated in Extended Data Table 1. We found that the errors in these measurements were 2.4 km s$^{-1}$ on average.

We derive a period using Fourier analysis. We used the Period04 software [34] to derive the system's period of 59.57±0.55 d, which is a peak with a signal to noise of 3.14, representing a 3σ significance, as shown in Extended Data Figure 2. The σ is described in depth in a paper describing the binary solution of the late-O star ι Ori [35], and we calculated the significance using both these same methods and the methods built into the Period04 [34] software yielding similar results, thus giving us 99% confidence in the period derived.

The measured positions were fit with an orbit using multiple fitting routines such as PHOEBE [36] and the python package BinaryStarSolver [37]. The results were similar and within the 1σ errors for both solutions and we report our best fit circular orbit in Table 1. Furthermore, we considered if eccentric orbits could better describe the system, but with the statistical methods for orbital solutions [15], we find that the system's orbit should be considered circular since the r.m.s. of the circular orbit is smaller than that of the eccentric solution. The eccentricity derived was 1σ of circular, with *e*=0.06±0.06, giving further confidence in our derivation of a circular orbit.

**Kinematics of the system in the Galactic Plane**

The system is located at Galactic coordinates (*l,b*)= (246.2, -0.6) and a distance of d=3.36±0.16 kpc from Gaia DR3 [28]. This places the system close to the Galactic plane $z$ = -36.0±1.7 pc in the Outer Perseus Arm of the Galaxy, and there are a number of massive OB-stars in its vicinity [11]. We used the coordinates, proper motion, and distance from Gaia DR3 together with the systemic velocity, γ, from Table 1 to find the components of peculiar velocity relative to its Local Standard of Rest. The estimates were made using the methods previously used to search for high velocity Be stars [39]. We find a peculiar tangential velocity of 5.8 ± 0.4 km s$^{-1}$ and a peculiar radial velocity of 14.2 ± 3.0 km s$^{-1}$, yielding a total peculiar space velocity of 15.3 ± 3.0 km s$^{-1}$. This is similar to but smaller than a previous estimate of the peculiar space velocity found to be 26.1 ± 15.7 km s$^{-1}$ [40]. The modest peculiar velocity is consistent with its location near its presumed birthplace in the plane, and it is also consistent with the relatively low runaway velocities of Be X-Ray binaries (average peculiar tangential velocity of $11 \pm 7$ km s$^{-1}$) [18,41].

# References

[24] Garmire, G. P. et al. "X-Ray and Gamma-Ray Telescopes and Instruments for Astronomy". Society of Photo-Optical Instrumentation Engineers (SPIE) Conference Series 4851, 28 (2003).

[25] https://cxc.harvard.edu/cal/ASPECT/celmon/

[26] https://cxc.cfa.harvard.edu/ciao/threads/reproject_aspect/

[27] Gaia Collaboration et al. "Gaia Data Release 2. Summary of the contents and survey properties". Astron. Astrophys. 616, 1 (2018). 1804.09365.

[28] Bailer-Jones, C. A. L., et al. "Estimating Distance from Parallaxes. IV. Distances to 1.33 Billion Stars in Gaia Data Release 2". Astron. J. 156, 58 (2018).

[29] Rivinius, Th., et al. "Classical Be stars. Rapidly rotating B stars with viscous Keplerian decretion disks". Astronomy and Astrophysics Review 21, 69 (2013).

[30] Watson, C. L., et al. "The International Variable Star Index (VSX)". Society for Astronomical Sciences Annual Symposium 25, 47 (2006).

[31] Samus', N. N., et al. "General catalogue of variable stars: Version GCVS 5.1". Astronomy Reports 61, 80 (2017).

[32] Tokovinin, A. et al. "CHIRON—A Fiber Fed Spectrometer for Precise Radial Velocities". Pub. Astron. Soc. Pac. 125, 1336 (2013). 1309.3971.

[33] Parades, L. et al. "The Solar Neighborhood XLVIII: Nine Giant Planets Orbiting Nearby K Dwarfs, and the CHIRON Spectrograph's Radial Velocity Performance". Astron. J. 162, 176 (2021).

[34] Lenz, P. & Breger, M. "Period04 User Guide". Comm. Asteroseismology 146, 53 (2005).

[35] Pablo, H. et al. "The most massive heartbeat: an in-depth analysis of  Orionis". Mon.

Not. R. Astron. Soc. 467, 2494 (2017).

[36] Prša, A. et al. "Physics Of Eclipsing Binaries. II. Toward the Increased Model Fidelity". Astrophys. J. Supp. Ser. 227, 29 (2016). 1609.08135.

[37] Barton, C. & Milson, N. BinaryStarSolver: Orbital elements of binary stars solver. Astrophysics Source Code Library 2012, 004 (2020).

[38] Bailer-Jones, C. A. L., et al. "Estimating Distances from Parallaxes. V. Geometric and Photogeometric Distances to 1.47 Billion Stars in Gaia Early Data Release 3". Astron. J. 161, 147 (2021).

[39] Berger, D. H. and Gies, D. R. "A Search for High-Velocity Be Stars". Astrophys. J. 555, 364 (2001).

[40] Boubert, D. and Evans, N. W. "galpy: A python Library for Galactic Dynamics". Astrophys. J. Supp. Ser. 216, 29 (2015).

[41] Bovy, J. "On the kinematics of a runaway Be star population". Mon. Not. R. Astron. Soc. 477, 5261 (2018).

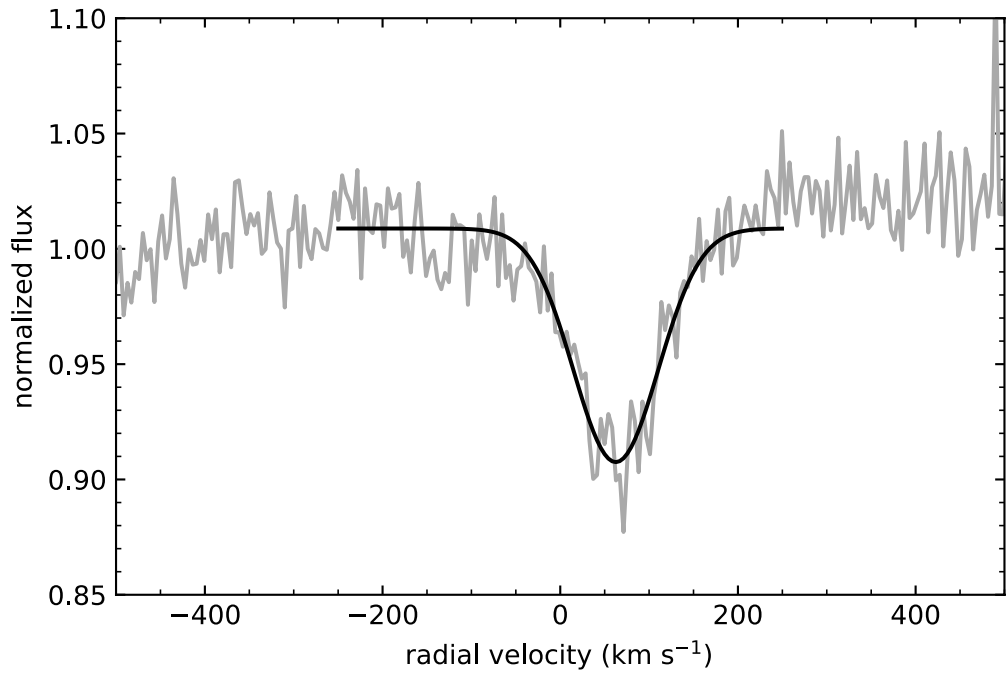

**Extended Data Figure 1.** The spectrum of CPD -29 2176 around the He II λ 4686 absorption line. Gaussian fits to the line were performed with the fitted minimum position representing our derived radial velocity for use in the orbital fits.

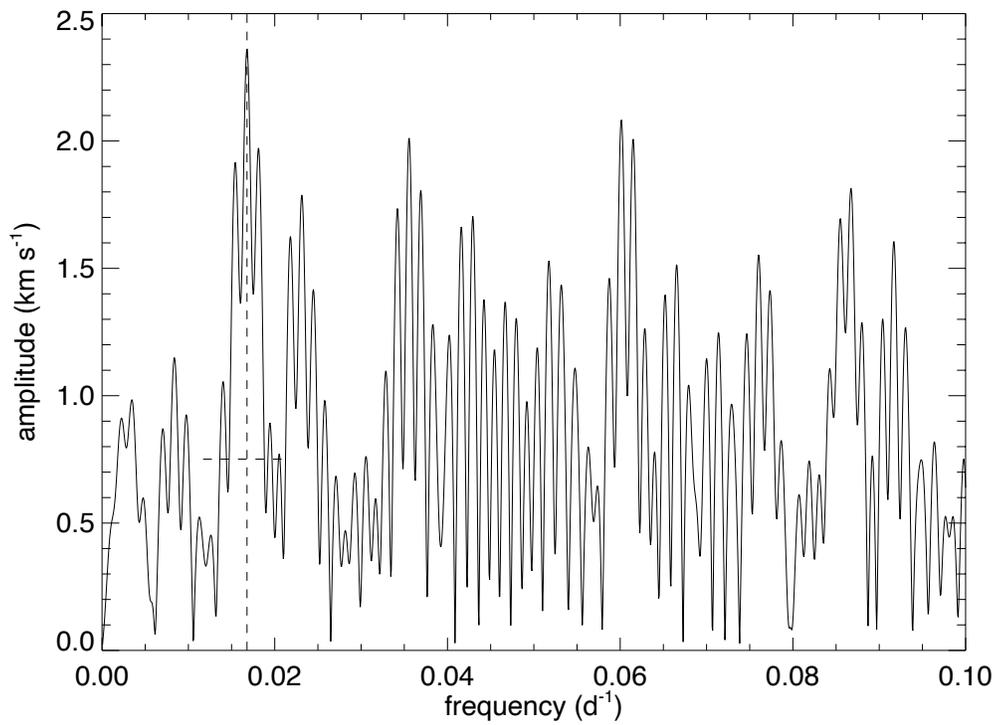

**Extended Data Figure 2.** The Fourier spectrum of the radial velocities shown in Extended Data Table 1. The peak at 0.016 d$^{-1}$ is our derived period for the system, and the noise level in the Fourier spectrum is denoted with a horizontal dashed line at that frequency, showing a 3σ significance for this peak.

| HJD -2,450,000 | RV (km s$^{-1}$) | Error (km s$^{-1}$) | S/N |
|---|---|---|---|
| 8392.8793 | 59.35 | 2.86 | 46 |
| 8393.8876 | 62.93 | 3.35 | 47 |
| 8410.7983 | 57.61 | 3.13 | 57 |
| 8440.8494 | 62.24 | 2.12 | 85 |
| 8452.8009 | 59.75 | 2.03 | 67 |
| 8466.7714 | 55.42 | 2.07 | 81 |
| 8494.6487 | 58.40 | 1.93 | 62 |
| 8509.6803 | 62.07 | 2.06 | 66 |
| 8543.5659 | 55.80 | 2.17 | 72 |
| 8551.6850 | 56.96 | 2.02 | 73 |
| 8558.6304 | 60.25 | 2.00 | 64 |
| 8564.5628 | 59.98 | 1.94 | 63 |
| 8582.6093 | 56.50 | 2.00 | 69 |
| 8597.5356 | 55.34 | 2.72 | 62 |
| 8624.5316 | 60.57 | 2.39 | 58 |
| 9197.7619 | 54.66 | 2.34 | 59 |
| 9208.7543 | 61.64 | 2.58 | 49 |
| 9209.7026 | 59.56 | 2.23 | 51 |
| 9219.7222 | 58.45 | 2.31 | 50 |
| 9224.6465 | 60.15 | 2.52 | 53 |
| 9228.7730 | 58.05 | 2.52 | 56 |
| 9234.6195 | 60.50 | 2.15 | 48 |
| 9240.6824 | 58.28 | 3.08 | 44 |

**Extended Data Table 1.** Measured Radial Velocities of He II 4686 absorption line.